\documentclass[aps,prl,twocolumn,superscriptaddress,showpacs,]{revtex4-1}

\usepackage{amsmath}
\usepackage{amssymb}
\usepackage{graphics}
\usepackage{graphicx}
\usepackage{epstopdf}
\usepackage{dcolumn}
\usepackage{bm}
\usepackage{xcolor}
\usepackage[marginal]{footmisc}
\usepackage[marginal]{footmisc}
\usepackage{flushend}
\usepackage{hyperref}
\usepackage{balance}
\hypersetup{hidelinks}

\newcommand{\beq}{\begin{equation}}
\newcommand{\eeq}{\end{equation}}
\newcommand{\bary}{\begin{array}}
\newcommand{\eary}{\end{array}}
\newcommand{\beqary}{\begin{eqnarray}}
\newcommand{\eeqary}{\end{eqnarray}}
\newcommand{\lgl}{{\langle}}
\newcommand{\rgl}{{\rangle}}

\newcommand{\Br}{{\bf r}}

\begin{document}

\title{Axial correlation revivals and number factorization with structured random waves}

\author{Xin Liu}
\email[]{These authors contributed equally to this work.}
\affiliation{Shandong Provincial Engineering and Technical Center of Light Manipulation \& Shandong Provincial Key Laboratory of Optics and Photonics Devices, School of Physics and Electronics, Shandong Normal University, Jinan 250014, China}
\affiliation{Collaborative Innovation Center of Light Manipulations and Applications, Shandong Normal University, Jinan 250358, China}
\author{Chunhao Liang}
\email[]{These authors contributed equally to this work.}
\affiliation{Shandong Provincial Engineering and Technical Center of Light Manipulation \& Shandong Provincial Key Laboratory of Optics and Photonics Devices, School of Physics and Electronics, Shandong Normal University, Jinan 250014, China}
\affiliation{Collaborative Innovation Center of Light Manipulations and Applications, Shandong Normal University, Jinan 250358, China}
\author{Yangjian Cai}
\email[]{yangjian\_cai@163.com}
\affiliation{Shandong Provincial Engineering and Technical Center of Light Manipulation \& Shandong Provincial Key Laboratory of Optics and Photonics Devices, School of Physics and Electronics, Shandong Normal University, Jinan 250014, China}
\affiliation{Collaborative Innovation Center of Light Manipulations and Applications, Shandong Normal University, Jinan 250358, China}
\author{Sergey A. Ponomarenko}
\email[]{serpo@dal.ca}
\affiliation{Department of Electrical and Computer Engineering, Dalhousie University, Halifax, Nova Scotia, B3J 2X4, Canada}
\affiliation{Department of Physics and Atmospheric Science, Dalhousie University, Halifax, Nova Scotia, B3H 4R2, Canada}

\date{\today}

\begin{abstract}
We advance a general theory of field correlation revivals of structured random wave packets, composed of superpositions of propagation-invariant modes, at pairs of planes transverse to the packet propagation direction. We derive an elegant analytical relation between the normalized intensity autocorrelation function of thus structured paraxial light fields at a pair of points on an optical axis of the system and a Gauss sum, thereby establishing a fundamental link between statistical optics and number theory. We propose and experimentally implement a simple, robust  analog random wave computer that can efficiently decompose numbers into prime factors. 
\end{abstract}

\maketitle

A spectacular self-imaging effect, first experimentally discovered by Talbot~\cite{Tb} and later theoretically explained by Lord Rayleigh~\cite{Ray}, occurs whenever a one-dimensional spatially periodic structure is illuminated by a freely propagating paraxial optical field. As was shown by Rayleigh~\cite{Ray}, the periodic field pattern formed by the structure undergoes periodic axial revivals at the distances that are multiple integers of the Talbot length, $z_T =2 d^2/\lambda$, where $d$ is a spatial period of the field pattern at the source and $\lambda$ is the wavelength of light. To date, the classic Talbot-like revivals have been extensively studied for coherent and random optical waves in free space~\cite{Win65,Mont67,Berry,Tlbrd1}, linear graded media~\cite{GRIN} and nonlinear optical systems~\cite{NLO}. Moreover, space-time duality~\cite{STduo} implies the existence of a temporal analogue of the spatial Talbot effect for a periodic train of optical pulses propagating in a dispersive fiber~\cite{Tlbtime1,Tlbtime2,Tlbtime3}. The spatial and temporal Talbot effects have found numerous applications to X-ray imaging~\cite{Xrays}, optical metrology and spectroscopy~\cite{Tlbrev}, as well as most recently to temporal cloaking~\cite{cloak}. Thanks to a mathematical analogy between the paraxial wave equation in optics and Schr\"{o}dinger equation in quantum mechanics, Talbot recurrences have been explored in atom optics~\cite{AO1,AO2,AO3}, Bose-Einstein condensates~\cite{BEC}, as well as in $C_{70}$ fullerene molecule~\cite{C70} and Rydberg wave packet~\cite{Ryd} interferometry. 

Lately, wave pattern revivals of structured fields have attracted interest~\cite{Gouy}, and novel manifestations of  spatial~\cite{PSA20}, temporal~\cite{PSA21a} as well as space-time~\cite{PSA21b} Talbot effects have been discovered for space-time wave packets for which spatial and temporal frequencies of individual monochromatic waves, composing the packets, are tightly coupled~\cite{ST-rev}. Further, aperiodic field correlations of an uncorrelated superposition of two Bessel beams at pairs of planes, transverse to the beam propagation direction in free space, have been shown to undergo perfect revivals over the Talbot distance~\cite{Tlbrd2}. 

 Here we develop a general theory of axial correlation revivals of structured random waves composed of superpositions of diffraction-free modes of any kind. Our theory enables us to derive a remarkably simple analytical relation between the normalized intensity autocorrelation function of structured paraxial random light fields at a pair of points on an optical axis of the system and an incomplete Gauss sum, thereby establishing a fundamental link between statistical optics and number theory. We employ the discovered link to advance and experimentally implement an efficient protocol to decompose even fairly large numbers into prime factors using random light. 
 
Number factorization plays a prominent role in network systems and cyber security~\cite{num-fac-bk}, as well as in optimization~\cite{opt1,opt2}. It has also become a crucial ingredient to a host of promising physics-based protocols for information encoding~\cite{HL-lsa}, optical encryption~\cite{LK-nc}, and all-optical machine learning~\cite{XL-sci}.  Moreover, intriguing prime number links have emerged to quantum ladder states of many-body systems~\cite{Gm-prl} and multifractality of light in  photonic arrays undergirded by algebraic number theory~\cite{Multi}.  

Although quantum algorithms have enabled seminal advances in number factorization~\cite{num-facNMR}, the application of quantum mechanics to this problem requires the implementation of a complex quantum Hamiltonian, such that factoring even relatively small numbers can run into unexpected difficulties~\cite{quant}. In addition, the entanglement of a large number of qubits is susceptible to pernicious decoherence effects~\cite{Quant}.  Hence, the alternatives have been sought that rely on the physics of classical superpositions of coherent waves ~\cite{Numfac1,Numfac2,Numfac3}. The latter, however, are very sensitive to external noise.  In contrast, our protocol involves classical random waves, which are robust against noise~\cite{SRL1,SRL2}, and its capacity is only limited by the pixel size of a spatial light modulator~\cite{Suppl}. 

We consider a random wave packet and an associated ensemble of scalar random fields $\{ U \}$. In the space-frequency representation, the field of each ensemble realization can be expressed as
\beq\label{wf}
		U(\Br, z,\omega)=\sum_{\nu} a_{\nu} \Psi_{\nu}(\Br,z,\omega),
			\eeq
where $z$ and $\Br$ are the axial coordinate along and radius vector in the plane transverse to the packet propagation direction, respectively. Further, $\{a_{\nu} \}$ is a set of uncorrelated random amplitudes which obey the second-order statistics as
\beq\label{stat}
		\lgl a_{\nu}^{\ast} a_{\nu^{\prime}}\rgl =\lambda_{\nu} \delta_{\nu\nu^{\prime}}.
			\eeq
Here the angle brackets represent ensemble averaging and $\{\lambda_{\nu}\geq 0\}$ specify powers of individual modes $\{\Psi_{\nu}\}$. 

We now specify to a particular class of spatial modes,
\beq\label{dfm}
		\Psi_{\nu}(\Br,z,\omega)=\psi_{\nu} (\Br,\omega)e^{i\beta_{\nu} (\omega)z},
			\eeq
and drop the irrelevant frequency variable $\omega$ hereafter. We can infer from Eq.~(\ref{dfm}) that each mode, characterized by a propagation constant $\beta_{\nu}$, defies diffraction. It follows that $\{\Psi_{\nu}\}$ are either modes of a chaotic multimode (optical/acoustical/matter) waveguide, or $\{\Psi_{\nu}\}$  belong to a special class of non-diffracting modes of the Helmholtz equation in free space. 

Next, we define the cross-spectral density of the ensemble at a pair of transverse planes $z_1=const$ and $z_2=const$ as 
\beq\label{W12-def}
		W(\Br_1,z_1; \Br_2, z_2)=\lgl U^{\ast}(\Br_1, z_1) U(\Br_2, z_2)\rgl.
			\eeq
On substituting from Eqs.~(\ref{wf}) through~(\ref{dfm}) into Eq.~(\ref{W12-def}), we can readily conclude that 
\beq\label{W12}
		W(\Br_1, \Br_2, \Delta z)= \sum_{\nu} \lambda_{\nu} \psi_{\nu}^{\ast}(\Br_1)\psi_{\nu}(\Br_2) e^{i\beta_{\nu} \Delta z},
			\eeq
where $\Delta z=z_2-z_1$. The analysis of Eq.~(\ref{W12}) reveals that the cross-spectral density of the field in any given transverse plane $z_1=z_2=const$ remains the same due to the propagation invariance of the modes. Yet, even uncorrelated modes manifest axial dynamics due to the interference of phasors $e^{i\beta_{\nu}\Delta z}$ when correlations in different transverse planes are examined. In particular, if the propagation constant is proportional to a polynomial in $\nu$ with integer coefficients,  $\beta_{\nu}=(2\pi/z_r)\sum_s c_s \nu^s$,  $c_s  \in \mathcal{Z},$ and $s \in \mathcal{N}$, $W(\Br_1, \Br_2, \Delta z)$ self-images over multiples of a characteristic revival distance $z_{r}$. This is a generic feature of axial correlations of the wave fields composed of discrete diffraction-free modes. 

\begin{figure}[t]
\centering
   \includegraphics[width=\linewidth]{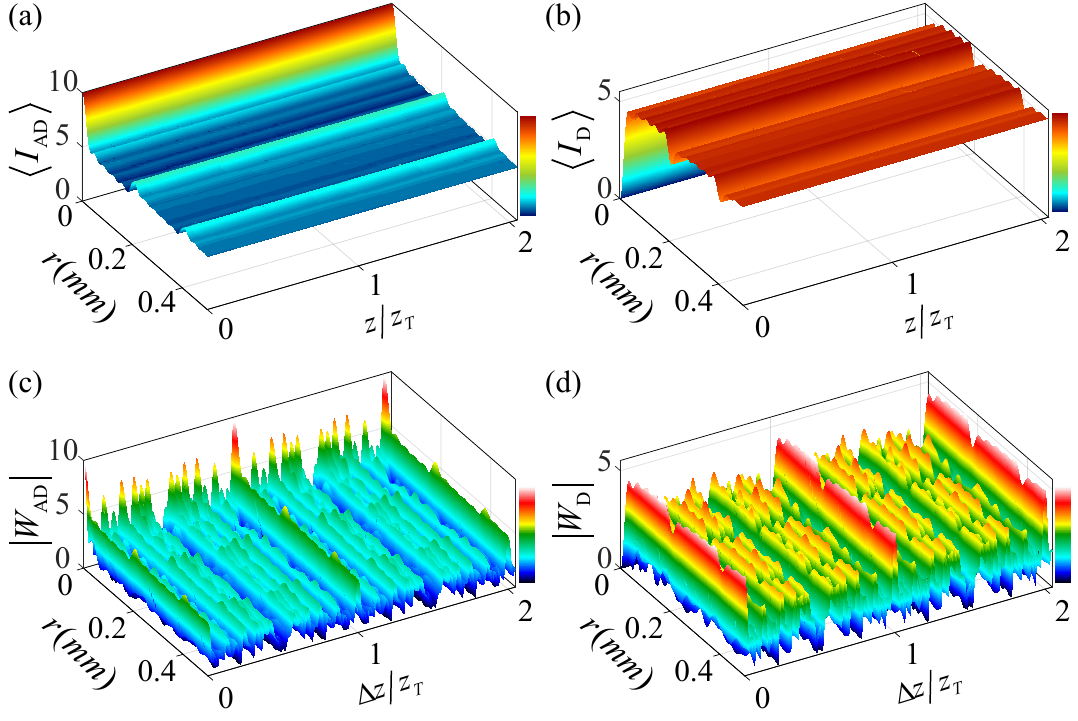}
   \caption{ Left/right: (Propagation-invariant) average intensity (a/b) and the magnitude of the cross-spectral density (c/d)  of an ensemble of antidark/dark beam superposition, evaluated at a pair of points with coordinates $(\Br,z_1)$ and $(\Br,z_2)$, $\Delta z=z_2-z_1$, using Eq.~(\ref{Wfn}) with $M=5$ and $d=0.4$ mm. At the multiples of the classic Talbot distance $z_T$, aperiodic DAD field correlations are perfectly reproduced. }
 \label{Tbt}
\end{figure}

Let us now focus on structured random sources generating uncorrelated superpositions of the so-called dark or antidark (DAD) diffraction-free beams of light, featuring dark notches or bright bumps against incoherent background ~\cite{DADt,DADexp}. Instructively, the DAD beams have been recently shown to maintain structural stability even in random media~\cite{DADturb}. 

We show (see Supplemental Material~\cite{Suppl} for details) that the cross-spectral density of the DAD beam superposition ensemble at a given radial position $r$ in any pair of transverse planes is given, up to an immaterial constant and overall phase factor, by the expression
\beq\label{Wfn}
		W_{\mathrm{DAD}}(\Br, \Delta z)\propto \sum_{m=1}^{M} [1\mp J_0 (4\pi m r/d )] e^{-2\pi i m^2 \Delta z/z_T},
			\eeq
where $J_0(x)$ is a Bessel function of the first kind and order zero, and $- (+)$ on the r.h.s of the equation corresponds to a superposition of dark (antidark) beams, respectively.
			
We display the average intensity and the magnitude of $W_{\mathrm{DAD}}(r, \Delta z)$ in Fig.~\ref{Tbt} for antidark (left panels) and dark (right panels) beams. We can observe in the figure that while the average intensity remains propagation-invariant,  the two-plane correlations exhibit, in general, intricate dynamics revealing perfect periodic revivals.  Specifically, we can infer from Eq.~(\ref{Wfn}) and observe in Fig.~\ref{Tbt}(c,d) that aperiodic DAD field correlations are perfectly reproduced at pairs of axial distances separated by multiple Talbot lengths. Further, two-point angular correlations of the ensemble of DAD beam superpositions show even more intriguing Talbot carpets, replete with full and fractional Talbot revivals~\cite{Suppl}. Instructively, we can introduce the second-order degree of coherence of the fields~\cite{MW} at a pair of points $(\Br,z_1)$ and $(\Br,z_1+\Delta z)$ as
\beq\label{mu-def}
		\mu_{\mathrm{DAD}}(\Br,\Delta z)=\frac{W_{\mathrm{DAD}}(\Br, \Delta z)}{W_{\mathrm{DAD}}(\Br,0)}.
			\eeq
It follows at once from Eqs.~(\ref{Wfn}) and~(\ref{mu-def}) that provided $\Delta z=n z_T$, $n \in \mathcal{N}$, the fields are perfectly coherent parallel to the optical axis at the transverse planes separated by multiple revival distances, $|\mu_{\mathrm{DAD}}(\Br,n z_T)|=1$. This is a statistical signature of Talbot correlation revivals which holds for random wave packets composed of any propagation-invariant modes $\{\Psi_{\nu} \}$. 

The presence of a quadratic exponential factor in Eq.~(\ref{Wfn}) hints at the possibility of employing correlations of structured random light fields to factor numbers. Indeed, we can establish a link between the normalized intensity autocorrelation function of the peak intensities of antidark (AD) beams $I_{\mathrm{AD}}$, evaluated at different positions along the optical axis, and an incomplete Gauss sum of number theory. To this end, we define the former as
\beq\label{g2}
		g_{\mathrm{AD}}^{(2)}(0,\Delta z)=\frac{\lgl I_{\mathrm{AD}}(0,z+\Delta z)I_{\mathrm{AD}}(0,z)\rgl}{\lgl I_{\mathrm{AD}}(0,z)\rgl^2}, 
			\eeq
and the latter as
\beq\label{Gs}
	\mathcal{G}_N^{(M)} (p)= \frac{1}{M} \sum_{m=1}^{M} e^{-2\pi i m^2 N/p},
			\eeq
and derive the following fundamental relation~ (see Supplemental Material~\cite{Suppl} for details):
\beq\label{num-opt}
		|\mathcal{G}_N^{(M)} (p)|^2= g_{\mathrm{AD}}^{(2)}(0,z_T N/p)-1.
			\eeq	
					
In Eqs.~(\ref{Gs}) and~(\ref{num-opt}), $N$ is a number that we seek to factor and $p$, $ p\in \mathcal{N}$, are trial prime integers that may or may not be factors of $N$. The left-hand side of Eq.~(\ref{num-opt}) equals to unity whenever $p$ is a factor of $N$ and it oscillates rapidly, taking on small values otherwise. Remarkably, we have revealed a link between a number theoretic quantity, the incomplete Gauss sum and a fundamental measurable quantity of statistical optics, the intensity autocorrelation function of light fields. It follows that we can sample $\mathcal{G}_N^{(M)}$ at discrete points by evaluating from experimental data (normalized) intensity-intensity correlations of light at pairs of points separated by the interval $\Delta z=z_T N/p$ along the optical axis of the system.

\begin{figure}[t]
\centering
   \includegraphics[width=\linewidth]{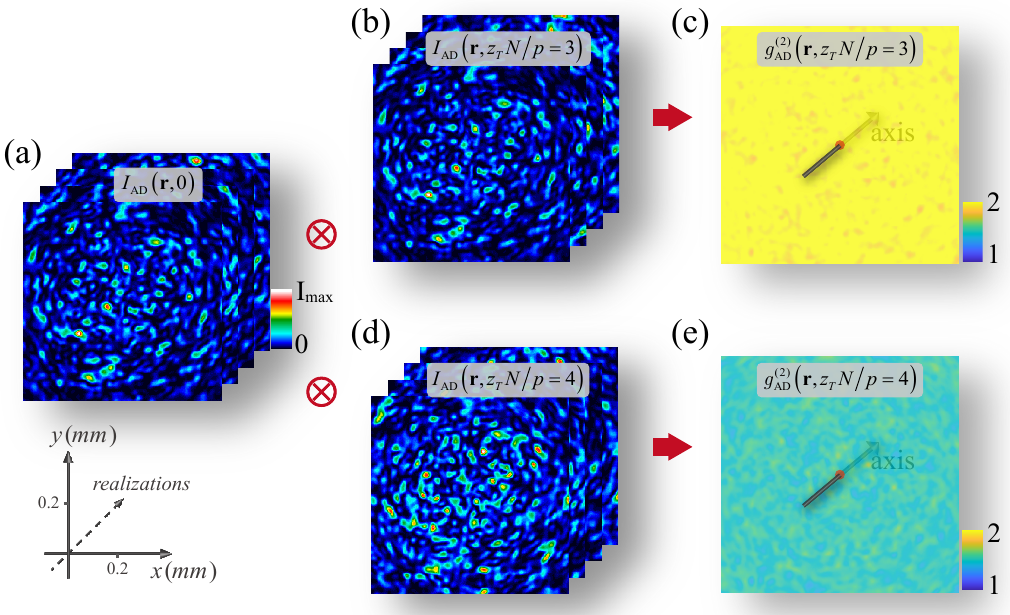}
   \caption{(a) (b) (d) Speckle patterns (instantaneous intensity) of the structured random light (AD beam with $M=5$ and $N=1155$) and (c) (e) the resulting normalized intensity autocorrelation functions. We choose $p=3$ in (a)(b)(c) and $p=4$ in (a)(d)(e), respectively. In the former case, $p$ is a factor of $N$, whereas it is not the case in the latter case. The normalized intensity autocorrelation functions $g_{\mathrm{AD}}^{(2)}(\Br,z_T N/p)$ in (c) and (e) are evaluated by averaging over an ensemble of 2000 speckle patterns with the aid of Eq. (S16) of the Supplemental Material~\cite{Suppl}, which reduces to Eq.~(\ref{g2}) on the optical axis. }
 \label{spec_corr}
 \end{figure}

Next, we validate the proposed protocol by carrying out a number factorization experiment. In our experiment, a collimated, linearly polarized quasi-monochromatic beam of carrier wavelength $\lambda=632.8$nm, emitted by a He-Ne laser, illuminates a phase-only spatial light modulator (SLM) (Meadowlark Optics, $1920\times1200$ 8\,$\mu$m$^2$ pixels). To structure random light beams, we encode the desired field distributions into holograms using complex amplitude encoding. The sought beams are then associated with the first diffracted order from the SLM; we refresh speckle patterns by refreshing the holograms, see Supplemental Material~\cite{Suppl} for further details. 

In Fig.~\ref{spec_corr} we present two typical ensemble representations (speckle patterns of instantaneous intensities) of light at the two axial distances corresponding to $p = 3$ (top row, $p$ being a factor of $N$) and $p = 4$ (bottom row) with $M = 5$ and $N = 1155$. We average over an ensemble of 2000 speckle patterns to evaluate the normalized intensity autocorrelation functions $g_{\mathrm{AD}}^{(2)}(\Br,z_T N/p)$  with the help of Eq. (S16) of the Supplemental Material~\cite{Suppl}. We can show analytically---see the Supplemental Material~\cite{Suppl}---that whenever $N=np$, $n \in \mathcal{N}$, $g_{\mathrm{AD}}^{(2)}(\Br,Nz_T/p )=2$ at any transverse location $\Br$, implying that $|\mathcal{G}_N^{(M)} (p)|^2=1$
in Eq.~(\ref{num-opt}). This conclusion is well supported by our experimental results of Fig.~\ref{spec_corr}(c) wherein slight deviations from the theory can be further suppressed by increasing the size of the ensemble. 
 
\begin{figure}[t!]
\centering
   \includegraphics[width=0.9\linewidth]{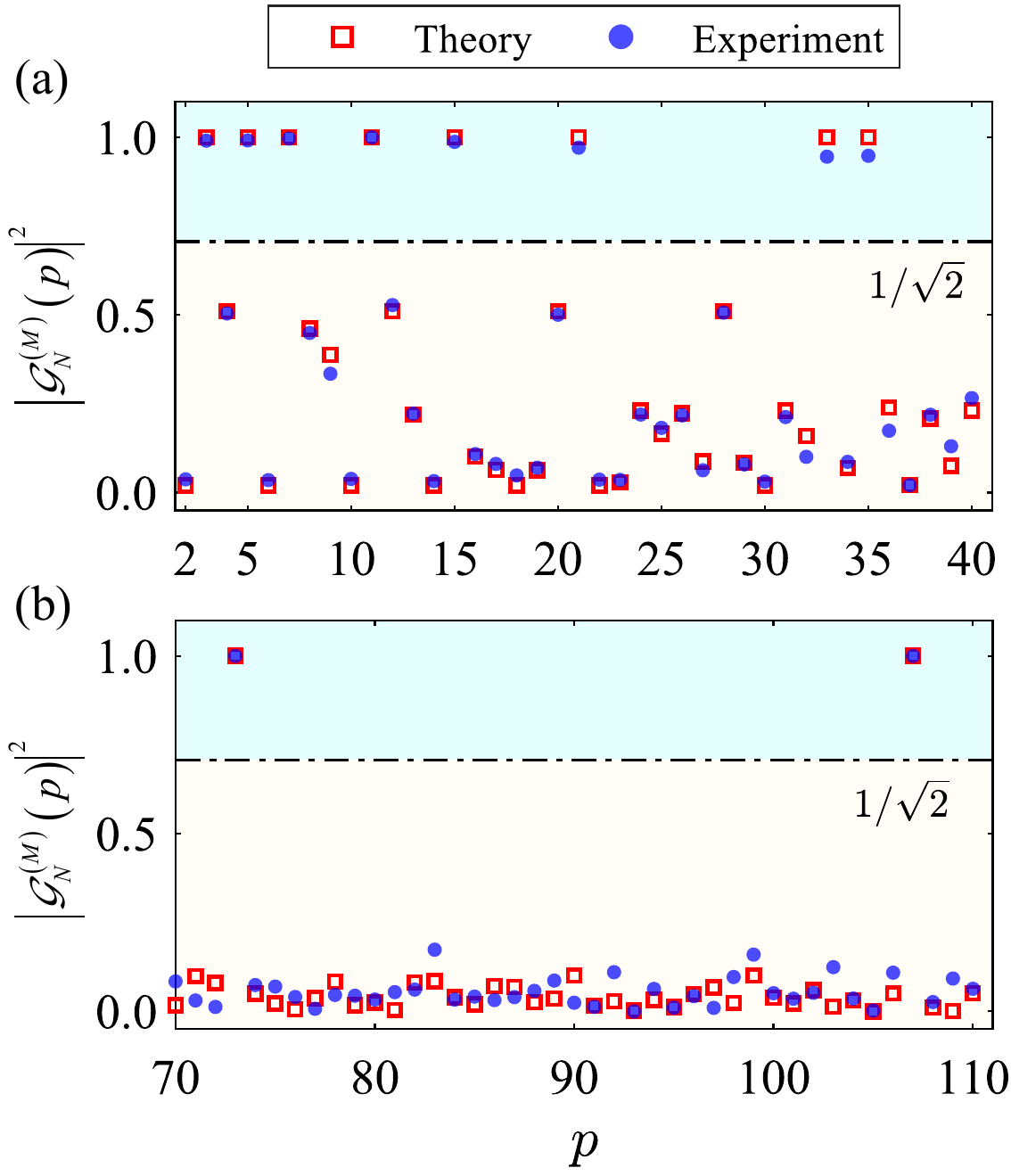}
   \caption{Experimental number factorization with structured random light beams. The relevant parameters are as follows: (a) $N = 1155 = 3\times5\times7\times11$ and $M = 5$, (b) $N = 570203= 73^2\times107$ and $M=20$. The theoretical and experimental results are obtained from Eqs.~(\ref{Gs}) and~(\ref{num-opt}), respectively. }
 \label{fac_num}
 \end{figure}

 To demonstrate the capability of our protocol to factor large numbers, we show theoretical and experimental results, corresponding to left- and right-hand sides of Eq.~(\ref{num-opt}), respectively,  as functions of $p$ in Fig.~\ref{fac_num}, which are marked by red squares (theory) and blue dots (experiment). In Fig.~\ref{fac_num}(a), a relatively small number to be factorized is comprised of four adjacent primes ($N=1155=3\times5\times7\times11$) whereas  Fig.~\ref{fac_num}(b) exhibits the factorization of a large number composed of prime factors that are far apart from one another: $N=570203=73^2\times107$. The results clearly demonstrate that $|\mathcal{G}_N^{(M)} (p)|^2$ attains unity (within the experimental accuracy) for all prime factors of $N$. Further, the factors and non-factors of $N$ are clearly discriminated by the threshold value of $1/\sqrt{2}$, as anticipated~\cite{Ms-njp}.

\begin{figure}[t!]
\centering
   \includegraphics[width=0.9\linewidth]{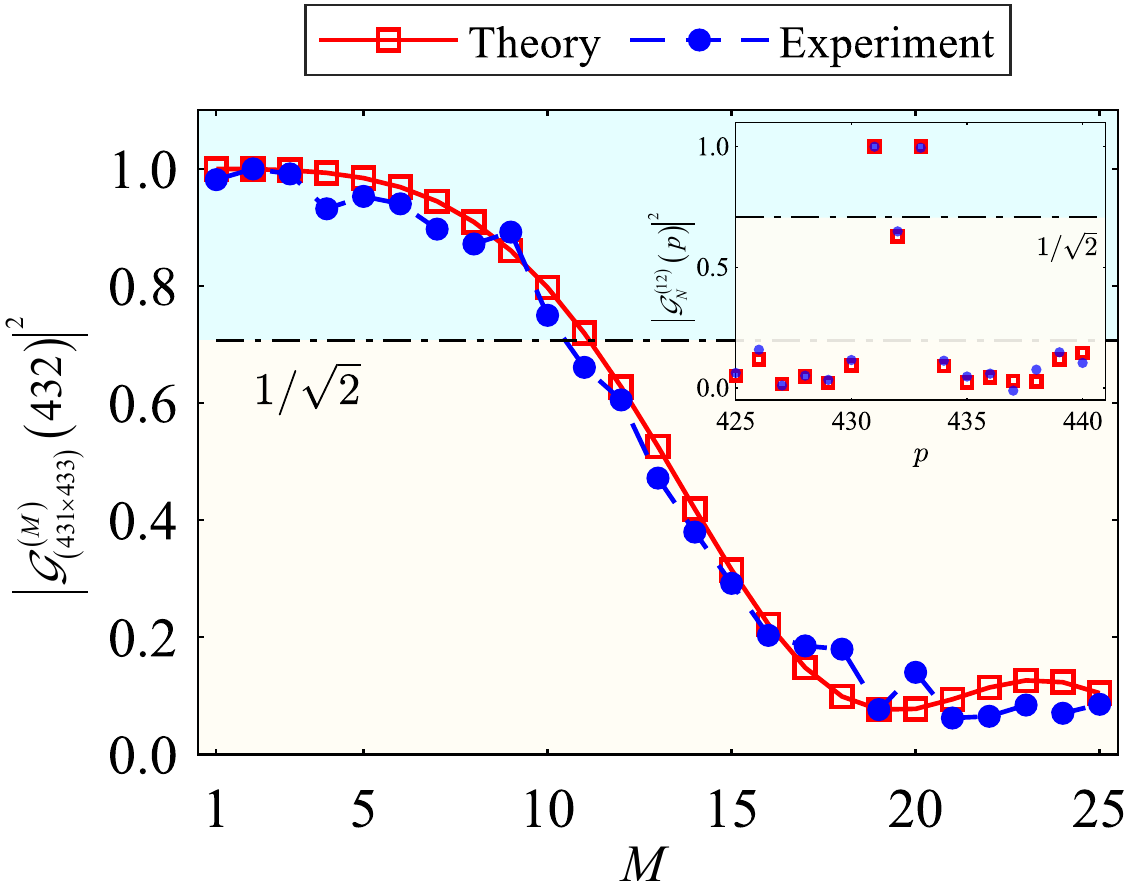}
   \caption{Experimental analysis of ghost factor $p=432$  suppression in the factorization of $N=186623=431\times433$ as function of the number of modes $M$. Inset: Experimental results of factoring $N=186623$ with $M=12$.}
 \label{Gho_num}
 \end{figure}

Next, to ensure the accuracy of number factorization, we demonstrate that our protocol is able to detect spurious factors, the ghost factors such that the magnitude of the incomplete Gauss sum can attain values close to unity. To suppress any ghost factor bellow the  threshold value, the magnitude of $M$ must be chosen judiciously. Previous work on number factorization with incomplete Gauss sums suggests the criterion $M\approx0.659\sqrt[4]{N}$ for a given $N$~\cite{Ms-njp}. To illustrate ghost number suppression in our protocol, we focus on a large number $N = 186623 =431\times433$, composed of two twin primes, and analyze the ghost factor 432. In Fig.~\ref{Gho_num}, we display the theoretical and experimental results for $|\mathcal{G}_{(431\times433)}^{(M)} (432)|^2$  as functions of $M$. We can infer from the figure that both curves fall off with $M$, exhibiting slight oscillations at their large $M$ tails. The curves intersect the horizontal line $|\mathcal{G}|^2=1/\sqrt{2}$ at $M=11.1$, implying that ghost factors are suppressed provided $M\geq12$. To verify this conclusion and impart intuition, we also plot $|\mathcal{G}_{(431\times433)}^{(12)}|^2$ as a function of $p$ evaluated for $M=12$  in the inset to the figure. We can readily conclude from the inset that the ghost factor 432 is clearly suppressed and only 431 and 433 are eligible prime factors of 186623. 

In conclusion, we have advanced a general theory of axial correlation revivals of structured random waves composed of superpositions of propagation-invariant modes. We have employed the developed theory to establish a fundamental link between statistical optical physics and number theory and have proposed and demonstrated a protocol to decompose numbers into prime factors using random light. It is worth noting that, to our knowledge, all previously reported classical physics inspired number factorization techniques relied on coherent superpositions of waves~\cite{Numfac1,Numfac2,Numfac3,Numfac4}. The strict control of wave phases was achieved by employing carefully controlled sequences of nuclear spins~\cite{Numfac1}, cold atoms~\cite{Numfac2}, or femtosecond optical pulses~\cite{Numfac3}, which, in turn, called for sophisticated experimental apparatuses~\cite{Numfac1,Numfac2,Numfac3} and cryogenic temperatures~\cite{Numfac2}. In contrast, our protocol requires only commercially available, table-top optical components and it employs cw random light, which is quite robust to source or environmental noise. It is then noteworthy that without making any attempt to optimize our procedure, we have been able to factor numbers as large as $N=570203$ which is, at least, on par with current achievements by NMR ($N=157573$)~\cite{Numfac1} and cold atom ($N=263193$)~\cite{Numfac2} based protocols and an order of magnitude larger than the greatest number ($N=19043)$ that has been factorized with coherent sequences of femtosecond pulses to date~\cite{Numfac3}. Moreover, the largest number amenable to factorization with coherent cw light beams has been mere 27~\cite{Numfac4}, so that our protocol enables the improvement of four orders of magnitude.  As a matter of fact, the upper bound on the number that can be decomposed into primes with our protocol is set by the pixel size of a commercial SLM employed in our experiment (see Supplemental Material for further details~\cite{Suppl}). As the emerging nanophotonics technology for structuring random light~\cite{meta-opt} matures, we fully anticipate further rapid advances in this direction. In addition, we are confident that similar number factorization protocols, undergirded by the fundamental principles that we have outlined above, can be developed to operate with noisy acoustical or matter waves to be structured with the aid of the appropriate metasurfaces~\cite{meta-acous,meta-mat}. 

The authors acknowledge financial support from the National Key Research and Development Program of China (2022YFA1404800, 2019YFA0705000), National Natural Science Foundation of China (12004220, 11974218, 12192254, 92250304), Regional Science and Technology Development Project of the Central Government (YDZX20203700001766), the China Postdoctoral Science Foundation (2022T150392), and the  Natural Sciences and Engineering Research Council of Canada (RGPIN-2018-05497).

\balance

\clearpage
\makeatletter
\renewcommand{\theequation}{S\arabic{equation}}
\renewcommand{\thefigure}{S\arabic{figure}}
\renewcommand{\thetable}{S\arabic{table}}
\pagebreak
\widetext
\begin{center}
\textbf{\large Supplemental Material: Axial correlation revivals and number factorization with structured random waves}
\end{center}

\setcounter{equation}{0}
\setcounter{figure}{0}
\setcounter{table}{0}
\setcounter{page}{1}
	
{\bf  S1. Derivation of the cross-spectral density of the ensembles of DAD beam superpositions at pairs of transverse planes.}---We recall that the cross-spectral density of any DAD beam ensemble in a single transverse plane can be expanded in terms of Bessel modes~\cite{DADt}. We generalize the latter to describe the ensembles of uncorrelated superpositions of DAD beams. Specifically, we consider the modes
	\beq\label{aux1}
		\Psi_{ml}(\Br,z)\propto J_{l} (k_m r) e^{il\phi} e^{i\sqrt{k^2 -k_m^2}z}. \tag{S1}
			\eeq
governed by a set of integer indices $\{\nu\}=\{m,l\}$, and we assume that
	\beq\label{aux2}
		k_m =2\pi m/d;  \hspace{0.5cm} 1\leq m\leq M, \tag{S2}
			\eeq
where $d$ is a characteristic transverse scale related to the width of the main lobe of each Bessel mode.  

We then take a (nonnegative) set of relative modal weights as
	\beq\label{aux3}
		\lambda_{ml} =1+ (-1)^l \alpha,  \hspace{0.5cm} |\alpha|\leq 1. \tag{S3}
		\eeq
Next, the paraxiality of the waves implies that $k_m\ll k=\omega/c$ for any $m$. It follows that 
	\beq\label{aux4}
		\sqrt{k^2-k_m^2}\simeq k-\frac{k_m^2}{2k}. \tag{S4}
			\eeq
On substituting from Eqs.~(\ref{aux1}) through~(\ref{aux4}) into Eq.~(5) of the main text, we arrive at the result
    \beq\label{aux5}
		W_{\mathrm{DAD}}(\Br_1, \Br_2, \Delta z)\propto e^{ik\Delta z}  \sum_{m=1}^{M}\sum_{l=-\infty}^{\infty} \lambda_{ml} J_l (k_m r_1) J_l (k_m r_2) e^{il (\phi_2-\phi_1)}
			\exp\left(-\frac{ik_m^2}{2k}\Delta z \right). \tag{S5}
			\eeq
Further, using the well-known addition theorem for Bessel functions~\cite{Gryz}
    \beq
		J_0 (k_m |\Br_1 \mp \Br_2|) =\sum_{l=-\infty}^{\infty} (\pm 1)^{l} J_l (k_m r_1) J_l (k_m r_2) e^{il (\phi_2-\phi_1)}, \tag{S6}
			\eeq
we can sum over the index $l$ to cast Eq.~(\ref{aux5}) into the form
	\beq\label{aux6}
		W_{\mathrm{DAD}}(\Br_1, \Br_2, \Delta z)\propto e^{ik\Delta z}  \sum_{m=1}^{M} [J_0 (k_m |\Br_1-\Br_2|) +\alpha J_0 (k_m |\Br_1+\Br_2|) ]\exp\left(-\frac{ik_m^2}{2k}\Delta z \right). \tag{S7}
			\eeq
			
At this stage, we can specify to correlations at the same point in the transverse plane, $\Br_1=\Br_2=\Br$, and assume the largest possible contrast index of $|\alpha|=1$ to reduce Eq.~(\ref{aux6}) to
	\beq\label{Wfn}
		W_{\mathrm{DAD}}(\Br, \Delta z)\propto e^{ik\Delta z}  \sum_{m=1}^{M} [1\mp  J_0 (4 \pi m r/d )] e^{-2\pi i m^2 \Delta z/z_T}, \tag{S8}
			\eeq
where the Talbot length $z_T$ is defined in the main text, and $- (+)$ on the r.h.s of the equation corresponds to a superposition of dark (antidark) beams, respectively. By the same token, the angular DAD correlations at pairs of transverse planes can be readily obtained from Eq.~(\ref{aux6}) by specializing to the case
$r_1=r_2$. The corresponding cross-spectral density reads
	\beq\label{aux7}
		W_{\mathrm{DAD}}(r, \Delta\phi, \Delta z)\propto e^{ik\Delta z}\sum_{m=1}^{M} [J_0 (2k_m r\sin\Delta\phi/2 )  \pm J_0 (2k_m r\cos\Delta\phi/2)]e^{-2\pi i m^2 \Delta z/z_T}, \tag{S9}
					\eeq
where $\Delta\phi=\phi_2-\phi_1$.
\begin{figure}[t]
\centering
   \includegraphics[width=0.8\linewidth]{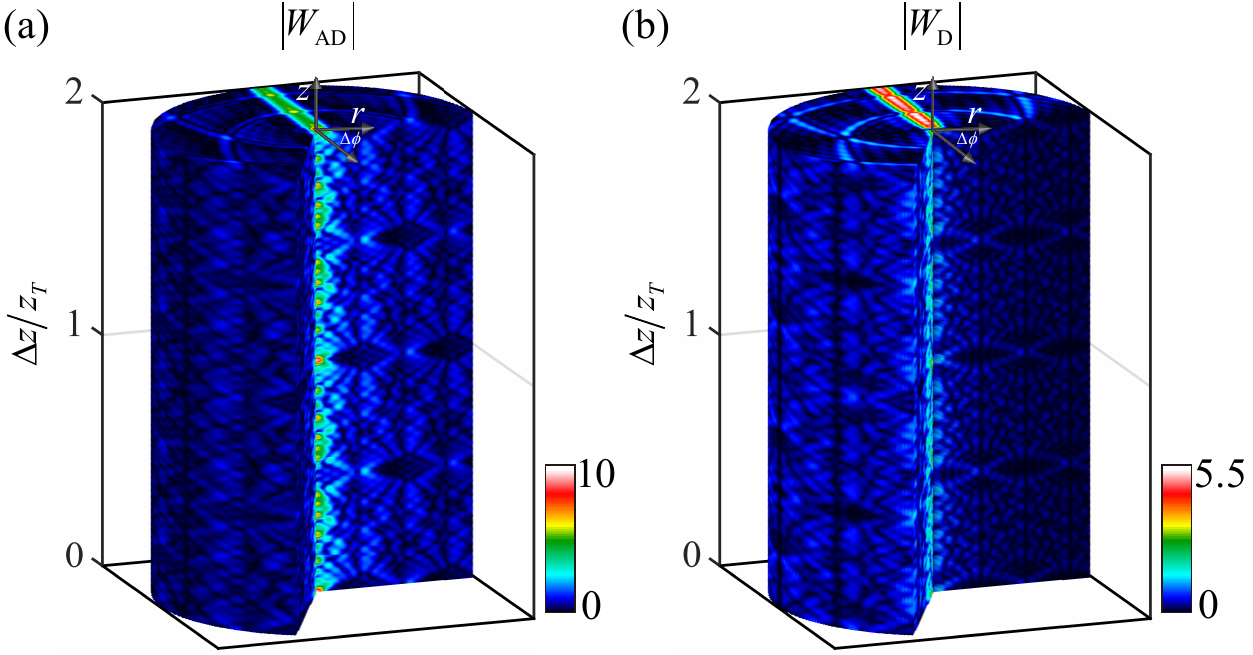}
   \caption{ Left/right: Magnitude of the cross-spectral density of an antidark/dark beam ensemble, evaluated at a pair of points with coordinates $(r,\phi_1, z_1)$, $(r, \phi_2, z_2)$, where $\Delta z=z_2-z_1$ and $\Delta\phi=\phi_2-\phi_1$ with the aid of Eq.~(\ref{aux7}) with $M=5$ and $d=0.4$ mm. }
 \label{Tb-phi}
\end{figure}
				
While the magnitude of the correlation function given by Eq.~(\ref{Wfn}) is displayed and described in the main text, we briefly discuss the azimuthal correlations here. First, we visualize azimuthal correlations of superpositions of DAD random fields in Fig.~1. We observe that the fields are strongly correlated for $\Delta\phi=0, \pi$ and only weakly correlated otherwise. Further, a rather complex Talbot carpet emerges as we scan along the radial coordinate $r$. In particular, we notice in Fig.~1a, that the antidark beam ensemble manifests a high contrast perfect revival over the Talbot length and radially shifted revival pattern over half the Talbot length; a similar scenario unfolds for the dark ensemble, albeit it has a relatively low off-axis contrast. At the same time, we notice a multitude of fractional revivals at the distances that are neither integer nor half-integer of $z_T$ from the source for either ensemble. 

{\bf S2. Derivation of the relation between normalized intensity autocorrelation function of an antidark beam ensemble and incomplete Gauss's sum.}---We start by introducing the second-order degree of coherence~\cite{MW} of an antidark beam superposition ensemble at a pair of points $(\Br,z)$ and $(\Br, z+\Delta z)$ by the expression
\beq\label{mu-def}
		\mu_{\mathrm{AD}}(\Br,\Delta z)=\frac{W_{\mathrm{AD}}(\Br, \Delta z)}{W_{\mathrm{AD}}(\Br,0)}, \tag{S10}
			\eeq
which can be viewed as the degree of coherence of light in a Young-type interference experiment with the pinholes located parallel to the optical axis of the system~\cite{PSA99}.  As random classical light obeys Gaussian statistics, the second-order degree of coherence at these points is related to the corresponding normalized intensity autocorrelation function $g^{(2)}$, defined as 
	\beq\label{aux8}
		g_{\mathrm{AD}}^{(2)}(\Br,\Delta z)=\frac{\lgl I_{\mathrm{AD}}(\Br, z+\Delta z)I_{\mathrm{AD}}(\Br, z)\rgl}{\lgl I_{\mathrm{AD}}(\Br, z)\rgl^2},  \tag{S11}
			\eeq
by the following Siegert relation~\cite{Good}:
	\beq\label{aux9}
		g_{\mathrm{AD}}^{(2)}(\Br, \Delta z) = 1+|\mu_{\mathrm{AD}}(\Br, \Delta z)|^2. \tag{S12}
			\eeq
Next, let us take the longitudinal distances such that
	\beq\label{aux10}
		\Delta z= Nz_T/p,   \hspace{0.5cm} p \in \mathcal{N}.  \tag{S13}
			\eeq
Here $p$ is a natural number that may or may not be a factor of a given odd number $N$ that we want to factorize. If we take now $N=np$, where $n \in \mathcal{N}$, implying that $p$ is a factor of $N$, we can infer at once from Eq~(\ref{aux10}) that $\Delta z=nz_T$. Therefore, all phasors on the right-hand side of Eq.~(\ref{Wfn}) are equal to unity. It then readily follows from Eqs~(\ref{Wfn}) and~(\ref{mu-def}) that 
	\beq\label{aux11}
		|\mu_{\mathrm{AD}}(\Br,\Delta z=nz_T)|=1, \tag{S14}
			\eeq
which, in turn, implies via Eq.~(\ref{aux9}) that 
	\beq\label{aux12}
		g_{\mathrm{AD}}^{(2)}(\Br, \Delta z=n z_T)=2. \tag{S15}
			\eeq  

Let us now focus on field correlations along the optical axis of the system. It follows at once from Eq.~(\ref{Wfn}), with the plus sign inside the brackets on the right-hand side, and Eq.~(\ref{mu-def}) that 
	\beq\label{aux13}
		\mu_{\mathrm{AD}} (0,\Delta z)=\frac{ e^{ik\Delta z}}{M}\,\sum_{m=1}^{M} e^{-2\pi i m^2 \Delta z/z_T}. \tag{S16}
			\eeq
Combining Eqs.~(\ref{aux10}) and~(\ref{aux13}), we obtain
	\beq\label{aux14}
		\mu_{\mathrm{AD}} (0,Nz_T/p) = e^{ik z_T N/p} \mathcal{G}_N^{(M)} (p), \tag{S17}
			\eeq
where we introduced an incomplete Gauss sum $\mathcal{G}_N^{(M)} (p)$ defined as~\cite{num,foot}
	\beq\label{aux15}
		\mathcal{G}_N^{(M)} (p)= \frac{1}{M} \sum_{m=1}^{M} e^{-2\pi i m^2 N/p}. \tag{S18}
			\eeq
Finally, it readily follows from Eqs.~(\ref{aux8}) and (\ref{aux9}), as well as from Eqs.~(\ref{aux14}) and (\ref{aux15}) that
	\beq\label{num-opt}
		|\mathcal{G}_N^{(M)} (p)|^2= g_{\mathrm{AD}}^{(2)}(0,z_T N/p)-1. \tag{S19}
			\eeq  

{\bf S3.  Experimental generation of structured random light beams.}---According to the tenets of statistical optics, any random light beam can be structured via a set of uncorrelated coherent modes~\cite{MW}. Specifically, each member of an ensemble of DAD beam superpositions with the cross-spectral density given by Eq. (\ref{aux5}) can be represented by the Karhunen-Lo\`{e}ve expansion~\cite{MW,Good} into coherent modes with random amplitudes. The field of such an ensemble realization can then be expressed as
	\beq\label{com-plex}
		U_{\mathrm{DAD}}(\Br,z) = \sum_{m=1}^{M} \sum_{l=-s}^{+s}a_{ml}J_{l}(k_mr)e^{il\phi}e^{ikz-ik_{m}^{2}z/2k}. \tag{S20}
			\eeq
Here $\{a_{ml}\}$ denotes a set of complex random amplitudes whose squared modulus and phase obey negative exponential and uniform distributions, respectively, which ensures Gaussian statistics of the ensemble. In principle, $s$ should span all integer values. We took $s=25$ in our experiment and have carefully checked that this subset of coherent modes nearly perfectly reproduces the ideal DAD beam intensity profile(s) following the methodology of~\cite{DADexp}.  

To generate the just described statistical ensemble in the laboratory, we need to encode complex electric fields of ensemble realizations into a phase-only SLM. The holograms that we load onto the SLM are obtained through the complex amplitude modulation encoding algorithm~\cite{Fb}. Specifically, the SLM phase is given by
     \beq\label{cp-modu}
           \Phi_{\mathrm{SLM}} = F_{E}(\Br,z)\sin\{\mathrm{Arg}[U_{\mathrm{DAD}}(\Br,z)]+2 \pi f_{x}x\}. \tag{S21}
                 \eeq
Here  “Arg”  stands for the principal value of the (random) phase of the field; $F_{E}$ is obtained via numerical inversion: $J_{1}(F_{E})=|U_{\mathrm{DAD}}|$, where $J_{1}$ denotes a Bessel function of the first kind and first order. Further, the phase shift $2\pi f_x x$ is imposed by a blazed grating, where $f_x$ denotes an inverse spatial period of the grating and determines the angular deviation in the $x$-direction of a light beam transmitted by the SLM. We can capture the intensities $I_{\mathrm{DAD}} = |U_{\mathrm{DAD}}(\Br,z)|^2$ of all ensemble realizations by refreshing the set of random amplitudes  $\{a_{ml}\}$ every time we generate a new realization. We then average over the ensemble of speckle patterns to evaluate the normalized intensity autocorrelation function of the structured random light fields  via Eq. (\ref{aux8}).

\begin{figure}[h!]
\centering
   \includegraphics[width=0.8\linewidth]{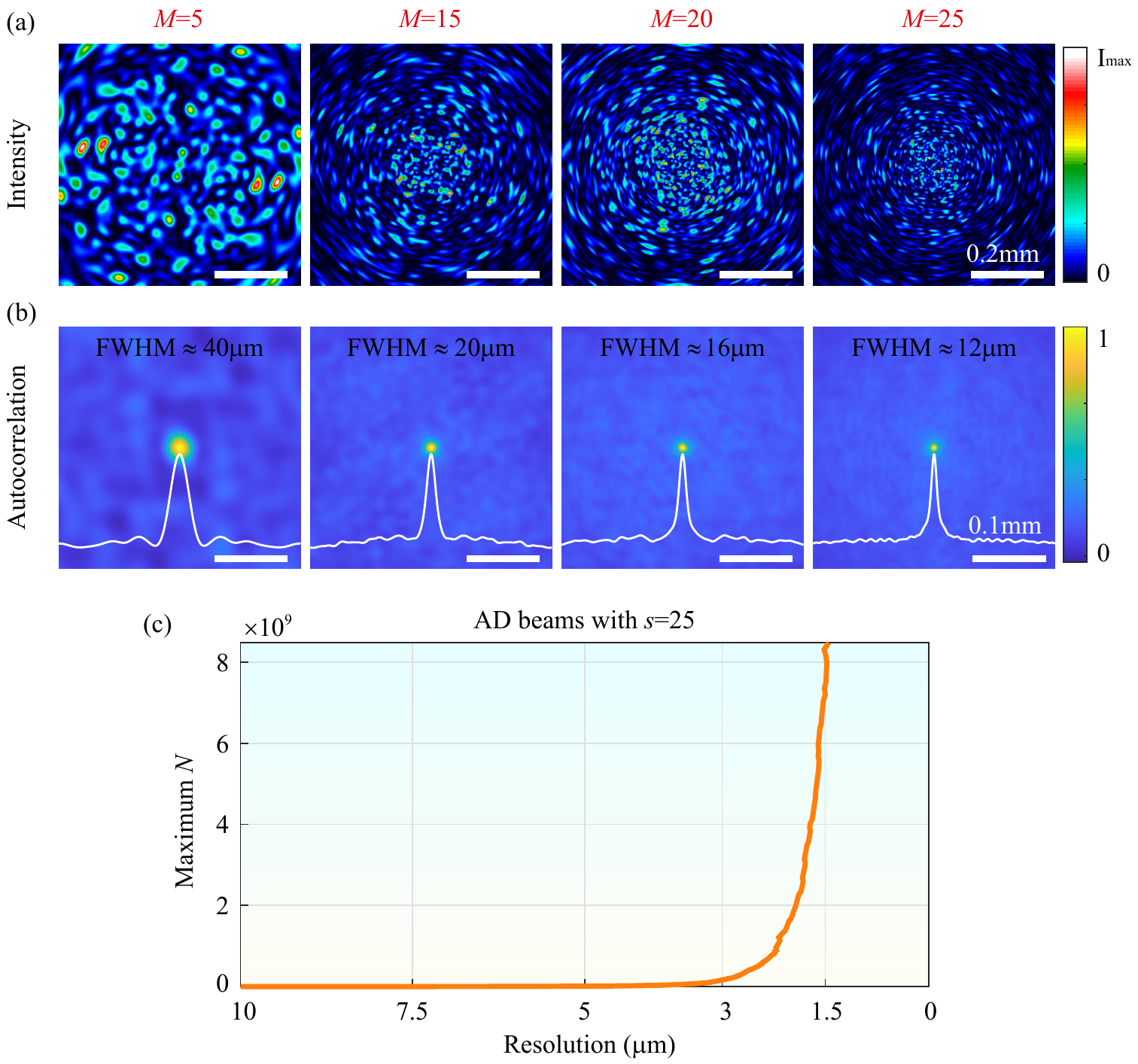}
   \caption{ Individual ensemble realization (speckle pattern) intensities (a) and the corresponding ensemble autocorrelation functions (b) of the structured random light beams (AD beams) with different number of modes $M$. (c) Numerical characterization of the maximum factorizable number $N$ of an AD beam as a function of spatial resolution of the light modulation device. Other parameters are as follows: $d=0.4$mm, $\alpha=1$ and $s=25$. }
 \label{Reso}
\end{figure}
{\bf S4. Upper bound for a number factorizable by our protocol.}---To accurately produce the structured random light beams (AD beams) in the lab, we must be able to resolve the fine structure of speckle patterns (ensemble realizations) in the experiment. According to Eq. (\ref{com-plex}), each realization can be viewed as a random superposition of Bessel beams of all orders up to $s$.  We anticipate that the size of a typical speckle grain within a random pattern shrinks as we increase either $M$ or $s$. Specifically, given $s=25$, we display generic speckle patterns of an antidark beam superposition as a function of $M$ in Fig.~\ref{Reso}(a). We can infer from the figure that the greater the magnitude of $M$, the finer the speckle grains, thus  imposing more stringent requirements on the spatial resolution of the available SLM. The Nyquist-Shannon sampling theorem dictates that each speckle grain be sampled by, at least, two pixels on the SLM screen~\cite{HL}. Therefore, the SLM pixel size  determines an ultimate upper bound on $M$, translating into an upper limit to the factorizable number by our protocol. Next, we follow the procedure outlined in Ref.~\cite{YP} to evaluate the speckle grain size, which is estimated by the full-width at half maximum (FWHM) of the autocorrelation function of the speckle ensemble. In Fig.~\ref{Reso} (b), we display the autocorrelation functions of the speckle ensembles corresponding to the same values of $M$ as those we selected to plot Fig.~\ref{Reso} (a).  The FWHM decreases from 40\,$\mu$m to 12\,$\mu$m as the parameter $M$ increases from 5 to 25. In our experiment, we employ an SLM from Meadowlark Optics, which has $1920\times1200$ pixels of the individual area of 8\,$\mu$m$^2$. It follows that the speckle grain size should be equal to, at least, 16\,$\mu$m to ensure resolution of individual speckle grains in our experiment. Therefore,  the parameter $M$ cannot exceed 20 [see the third column in Fig.~\ref{Reso} (a) and (b)]. Hence, the criterion $M\approx0.659\sqrt[4]{N}$~\cite{MS} implies the upper limit to a factorizable number by our protocol to be around $8.5\times10^5$. Of course, this is just operational, not principle limitation and we can factor larger numbers if we employ either higher resolution (smaller pixel size) SLMs, or other light modulators such as metasurfaces. To illustrate the dependence of the maximum facorizable number in our protocol on the light modulator resolution, we present in Fig.~\ref{Reso} (c)  the numerical characterization of the maximum factorizable number $N$ as a function of spatial resolution. We can infer from the figure that factoring numbers as large as $10^{10}$ with our protocol is quite within the limits of the existing technology.


\begin{thebibliography}{99}
\bibitem{foot} 
	The incomplete Gauss sum is sometimes alternatively defined as 
    $\tilde{\mathcal{G}}_N^{(M)} (p)=\frac{1}{M+1} \sum_{m=0}^{M} e^{-2\pi i m^2 N/p}$. The $m=0$ term is a trivial constant, though, which does not affect the factorization protocol. The two definitions are related viz., $\tilde{\mathcal{G}}_N^{(M)} (p)=\textstyle\frac{1}{M+1}\left[1+M \mathcal{G}_N^{(M)} (p)\right]$. It follows that for any $p$ that is a factor of $N$, both definitions yield unity, while both wildly oscillate for the other $p$'s, yielding small numbers.  Further, as $M$ increases, which is required for accurate factorization of large $N$'s, the magnitudes of the two sums become virtually indistinguishable.
\end{thebibliography}
\end{document}